%% file: main.tex
\documentclass{article}
\usepackage[utf8]{inputenc}
\usepackage[a4paper,top=3cm,bottom=2cm,left=2.5cm,right=2.5cm]{geometry}

\usepackage{authblk}

\usepackage{algorithm}
\usepackage{color}  
\usepackage[noend]{algpseudocode}
\usepackage{float}
\usepackage{comment}
\usepackage[numbers]{natbib}
\setcitestyle{numbers,open={[},close={]}}
\usepackage{amsmath}
\usepackage{amsthm}
\usepackage{booktabs}
\usepackage{tikz}
\usepackage{todonotes}
\usetikzlibrary{positioning,chains,fit,shapes,calc}
\definecolor{myblue}{RGB}{80,80,160}
\definecolor{mygreen}{RGB}{80,160,80}

\newcommand{\plus}{\raisebox{.4\height}{\scalebox{.6}{\textbf{+}}}}

\newcommand{\cala}{\mathcal{A}}

\newcommand{\caln}{\mathcal{N}}

\newcommand{\OPT}{\ensuremath{\mathit{OPT}}}

\usepackage
{hyperref}

  \title{Local-Search Based Heuristics for Advertisement Scheduling}
 
\author[1]{Mauro R. C. da Silva}
\author[1]{Rafael C. S. Schouery}
\affil[1]{Institute of Computing\\ University of Campinas}

\affil[ ]{\textit {maurorcsc@gmail.com and rafael@unicamp.br}}

\begin{document}

\maketitle

\section*{Abstract}
In the MAXSPACE problem, given a set of ads $\cala$, one wants to place a subset $\cala'\subseteq\cala$ into $K$ slots $B_1, \dots, B_K$ of size $L$. Each ad~$A_i \in \cala$ has \textit{size}~$s_i$ and \textit{frequency}~$w_i$. A schedule is feasible if the total size of ads in any slot is at most $L$, and each ad $A_i \in \cala'$ appears in exactly $w_i$ slots. The goal is to find a feasible schedule that maximizes the space occupied in all slots. We introduce MAXSPACE-RDWV, a MAXSPACE generalization with release dates, deadlines, variable frequency, and generalized profit. In MAXSPACE-RDWV each ad~$A_i$ has a release date~${r_i \geq 1}$, a deadline~${d_i \geq r_i}$, a profit~$v_i$ that may not be related with~$s_i$ and lower and upper bounds~$w^{min}_{i}$ and~$w^{max}_{i}$ for frequency. In this problem, an ad may only appear in a slot~$B_j$ with~${r_i \le j \le d_i}$, and the goal is to find a feasible schedule that maximizes the sum of values of scheduled ads. This paper presents some algorithms based on meta-heuristics GRASP, VNS, and Tabu Search for MAX\-SPACE and MAXSPACE-RDWV\@. We compare our proposed algorithms with~Hybrid-GA proposed by~\citet{kumar2006scheduling}. We also create a version of Hybrid-GA for MAXSPACE-RDWV and compare it with our meta-heuristics. Some meta-heuristics, such as VNS and GRASP$\plus$VNS, have better results than Hybrid-GA for both problems. In our heuristics, we apply a technique that alternates between maximizing and minimizing the filling of slots to obtain better solutions. We also applied a data structure called BIT to the neighborhood computation in MAXSPACE-RDWV and showed that this enabled ours algorithms to run more iterations.

\textbf{Keywords} packing; scheduling; advertisements; local-search; heuristics

\section{Introduction}
\input{introducao.tex}

\section{Literature Review}
\input{revisao.tex}

\section{Heuristics}
\input{heuristicas.tex}

\section{Computational Experiments}
\input{experimentos.tex}

\section{Results}
\label{sec:res}
\input{resultados.tex}


\section*{Acknowledgments}

This project was supported by S\~ao Paulo Research Foundation~(FAPESP) grants \mbox{\#2015/11937-9}, \mbox{\#2016/23552-7}, \mbox{\#2017/21297-2} and \mbox{\#2020/13162-2}, and National Council for Scientific and Technological Development~(CNPq) grants \mbox{\#425340/2016-3}, \mbox{\#308689/2017-8} and \mbox{\#425806/2018-9}.

\bibliography{main}
\end{document}

%% file: introducao.tex
The revenue from web advertising grew considerably in the 21st century. In 2019, the total revenue was~US\$124{.}6 billion, an increase of~15{.}9\% from the previous year. It is estimated that web advertising comprised~43{.}4\% of all advertising spending, overtaking television advertising~\citep{silverman2020iab}.

Many websites (such as Google, Yahoo!, Facebook, and others) offer free services while displaying advertisements, or simply \textit{ads}, to users. Often, each website has a single strip of fixed height, which is reserved for scheduling ads, and the set of displayed ads changes on a time basis. For such websites, the advertisement is the main source of revenue. Thus, it is important to find the best way to dispose the ads in the available time and space while maximizing the revenue~\citep{kumar2015optimization}.

 Websites like Facebook and Mercado Livre (a large Latin American marketplace) use banners to display advertisements while users browse. Google displays ads sold through Google Ad Words in its search results within a limited area, in which ads are in text format and have sizes that vary according to the price. In 2019, ads in banners comprised~31\% of internet advertising (considering banners and mobile platforms), which represents a revenue of~US\$38{.}62 billion~\citep{silverman2020iab}. Web advertising has created a multi-billionaire industry where algorithms for scheduling advertisements play a significant role.

In this paper, we consider the class of Advertisement Scheduling Problems introduced by~\citet{adler2002scheduling}, in which, given a set~${\cala = \{A_1, A_2, \dots, A_n\}}$ of advertisements, the goal is to schedule a subset~${\cala' \subseteq \cala}$ into a banner in~$K$ equal time-intervals. The set of ads scheduled to a particular time interval~$j$,~${1 \le j \le K}$, is represented by a set of ads~${B_j \subseteq \cala'}$, which is called a~\textit{slot}. Each ad~$A_i$ has a \textit{size}~$s_i$ and a \textit{frequency}~$w_i$ associated with it. Size~$s_i$ represents the amount of space~$A_i$ occupies in a slot, and frequency~${w_i \leq K}$ represents the number of slots which should contain a copy of~$A_i$. An ad~$A_i$ can be displayed at most once in a slot, and~$A_i$ is said to be \textit{scheduled} if exactly~$w_i$ copies of~$A_i$ appear in slots~\citep{adler2002scheduling, dawande2003performance}.

The main problems in this class are MINSPACE and MAXSPACE\@. In MINSPACE, all ads need to be scheduled in the slots, and the goal is to minimize the fullness of the fullest slot. In MAXSPACE, the focus of this paper, an upper bound~$L$ is specified, which represents the size of each slot. A feasible solution for this problem is a schedule of a subset~${\cala' \subseteq \cala}$ into slots~${B_1, B_2, \dots, B_K}$, such that each~${A_i \in \cala'}$ is scheduled and the fullness of any slot does not exceed the upper bound~$L$, that is, for each slot~$B_j$,~${\sum_{A_i \in B_j}{s_i}\leq L}$. The goal of MAXSPACE is to maximize the fullness of the slots, defined by~$\sum_{A_i \in \cala'}{s_i w_i}$. Both of these problems are strongly NP-hard~\citep{adler2002scheduling, dawande2003performance}.

To illustrate both problems, consider the ads in Table~\ref{tab:prop}.

\begin{table}[ht]
\centering
\caption{Example with~$7$ ads.\label{tab:prop}}
\begin{tabular*}{\textwidth}{c @{\extracolsep{\fill}}ccccccc}
\toprule
\textbf{$A_i$} & $A_1$ & $A_2$ & $A_3$ & $A_4$ & $A_5$ & $A_6$ & $A_7$ \\ \midrule
\textbf{$s_i$} & 6     & 4     & 2     & 3     & 1     & 1     & 5     \\
\textbf{$w_i$} & 3     & 2     & 1     & 2     & 1     & 1     & 1     \\ \bottomrule
\end{tabular*}
\end{table}

\begin{figure}[H]
  \centering
      \includegraphics[width=.75\textwidth]{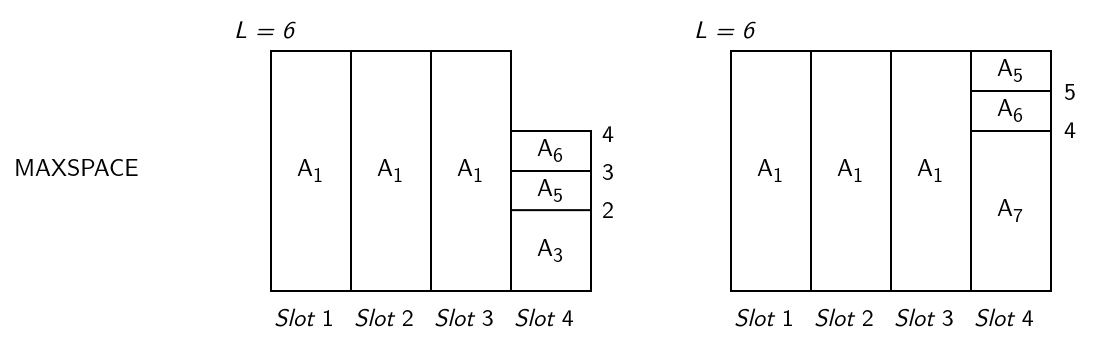}
  \caption[Example of solutions to MAXSPACE.]{Example of solutions to MAXSPACE using ads of Table~\ref{tab:prop} with~${L = 6}$ and~${K = 4}$. In~(a), we have a feasible solution. In~(b), we have an optimal solution~\citep{dawande2003performance}.\label{fig:max}}
\end{figure}

\begin{figure}[H]
  \centering
      \includegraphics[width=.75\textwidth]{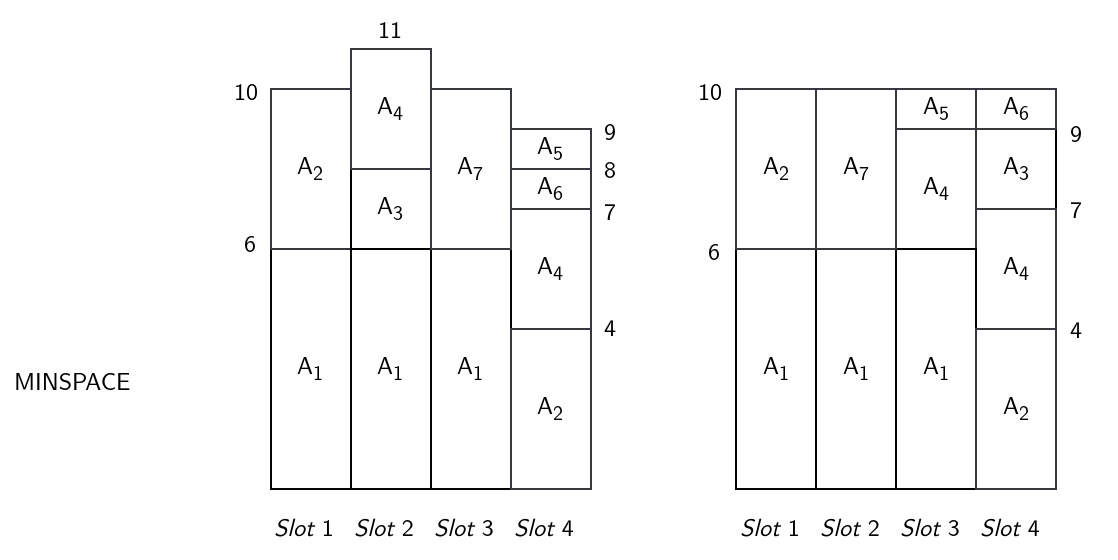}
  \caption[Example of solutions to MINSPACE.]{Example of solutions to MINSPACE using ads of Table~\ref{tab:prop} with~${K = 4}$. In~(a) we have a feasible solution with value~$11$. In~(b) we have an optimal solution with value~$10$~\citep{dawande2003performance}.\label{fig:min}}
\end{figure}

In Figures~\ref{fig:max} and~\ref{fig:min} we present solutions to MAXSPACE and MINSPACE, respectively, with the ads of Table~\ref{tab:prop}.

\citet{dawande2003performance} provide the following Integer Linear Programming formulation for MAXSPACE\@. Let~$x_{ij}$ be an integer variable that has value~$1$ if ad~$A_i$ was added to slot~$B_j$ and has value~$0$ otherwise, and let~$y_{i}$ be an integer variable that has value~$1$ when ad~$A_i$ was added into any slot and~$y_{i}$ is~$0$ otherwise. Then, we can formulate MAXSPACE as follows.
\begin{equation*}
\begin{array}{lll}
\max\displaystyle \sum\limits^{K}_{j=1}\sum\limits^{n}_{i=1}s_i x_{ij}&\\
\text{subject to}&\displaystyle \sum\limits_{i = 1}^{n} s_i x_{ij} \leq L,  &j=1, 2,\dots, K\\
                 & \displaystyle \sum\limits_{j = 1}^{K} x_{ij} = w_i y_i,  &i=1, 2,\dots, n\\
                 & \displaystyle x_{ij} \in \{0, 1\}, y_{i} \in \{0, 1\},  &i=1, 2,\dots, n, j=1, 2,\dots, K\\
                 
\end{array}
\end{equation*}
The first set of constraints ensures that the fullness of each slot must be at most~$L$, and the second set of constraints ensures that each scheduled ad~$A_i$ must be added to exactly~$w_i$ slots.

\citet{dawande2003performance} also provide the following Integer Linear Programming formulation for MINSPACE\@. Again, let~$x_{ij}$ be an integer variable that has value~$1$ if ad~$A_i$ was added to slot~$B_j$ and has value~$0$ otherwise. The formulation is as follows.
\begin{equation*}
\begin{array}{lll}
\min\displaystyle F&\\ 
\text{subject to}&\displaystyle F \geq \sum\limits_{i = 1}^{n} s_i x_{ij},  &j=1, 2,\dots, K\\
                 & \displaystyle \sum\limits_{j = 1}^{K} x_{ij} = w_i,  &i=1, 2,\dots, n\\
                 & \displaystyle x_{ij} \in \{0, 1\},  &i=1, 2,\dots, n, j=1, 2,\dots, K\\
                 
\end{array}
\end{equation*}
The first set of constraints together with the objective minimize the height of the schedule, and the second set of constraints ensures that each scheduled ad~$A_i$ must be added to exactly~$w_i$ slots.

\subsection{Proposed Problem}

The original MAXSPACE problem considers the value of an ad as the space it occupies (its size multiplied by the number of times it appears). In practice, the value of an ad can be influenced by other factors, such as the (expected) number of clicks the ad generates to the advertiser~\citep{briggs1997advertising}.

The time interval relative to each slot in the scheduling of advertising can represent minutes, seconds, or long periods, such as days and weeks. Often, one considers the idea of \textit{release dates} and \textit{deadlines}. An ad has a release date that indicates the beginning of its advertising campaign. Analogously, the deadline of an ad indicates the end of its advertising campaign. For example, ads for Christmas must be scheduled before December,~25th. Thus, ads with small deadlines must be prioritized while scheduling.

The number of times the ad appears can also be influenced by other factors, such as the advertiser's budget.
A variant that can be interesting in practice considers that each ad has a budget (instead of a frequency), which is reduced when some copy is placed.

With these observations in mind, we introduce MAXSPACE-RDWV, a MAXSPACE variant with release dates, deadlines, variable frequency, and generalized profit. In MAXSPACE-RDWV, each ad~$A_i$ has a release date~${r_i \geq 1}$, a deadline~${d_i \geq r_i}$, a profit~$v_i$ that may not be related with~$s_i$ and
lower and upper bounds~$w^{min}_{i}$ and~$w^{max}_{i}$ for frequency.
The release date of ad~$A_i$ represents the first slot where a copy of~$A_i$ can be scheduled; that is, a copy of~$A_i$ cannot be scheduled in a slot~$B_j$ with~${j < r_i}$. Similarly, the deadline of an ad~$A_i$ represents the last slot where we can schedule a copy of~$A_i$, thus~$A_i$ cannot be scheduled in a slot~$B_j$ with~${j > d_i}$. The goal is to find a feasible schedule that maximizes the sum of values of scheduled ads.
Note that MAXSPACE is a particular case of MAXSPACE-RDWV in which each ad~$A_i$ has~${w^{min}_{i} = w^{max}_{i} = w_i}$,~${v_i = s_i}$,~${r_i = 1}$ and~${d_i = K}$.

We provide the following Integer Linear Programming formulation for MAXSPACE-RDWV\@. Let~$x_{ij}$ be an integer variable that has value~$1$ if ad~$A_i$ was added to slot~$B_j$ and has value~$0$ otherwise, and let~$y_{i}$ be an integer variable that has value~$1$ when ad~$A_i$ was added into any slot and~$y_{i}$ is~$0$ otherwise. Then, we can adapt the formulation of~\citet{dawande2003performance} as follows.
\begin{equation*}
\begin{array}{lll}
\max\displaystyle \sum\limits^{K}_{j=1}\sum\limits^{n}_{i=1}v_i x_{ij}&\\
\text{subject to}&\displaystyle \sum\limits_{i = 1}^{n} s_i x_{ij} \leq L,  &j=1, 2,\dots, K\\
                 & \displaystyle \sum\limits_{j = r_i}^{d_i} x_{ij} = w_i y_i,  &i=1, 2,\dots, n\\
                 & \displaystyle x_{ij} = 0,  &i=1, 2,\dots, n, j \not \in \{r_i, r_i + 1, \dots, d_i\}\\
                 & \displaystyle x_{ij} \in \{0, 1\}, y_{i} \in \{0, 1\},  &i=1, 2,\dots, n, j=1, 2,\dots, K\\
                 
\end{array}
\end{equation*}
The first set of constraints ensures that the fullness of each slot must be at most~$L$, the second set of constraints ensures that each scheduled ad~$A_i$ must be added to exactly~$w_i$ slots and the third set of constraints ensures that no copies will be added before the release date or after the deadline.

As the above ILP formulation is unable to solve large instances of the problem, this paper presents some algorithms based on the meta-heuristics Greedy Randomized Adaptive Search Procedure~(GRASP), Variable Neighborhood Search~(VNS), Local Search, and Tabu Search for MAXSPACE and MAXSPACE-RDWV\@. 

To obtain better solutions, we alternate, depending on the iteration, between maximizing and minimizing a secondary objective function that computes the sum of the square of empty space in the slots. This allows to alternate between packing more copies of already packed ads and adding unpacked ads to the solution.

We also use a data structure to check a necessary condition to see if an unpacked ad can be added to the current solution. Even though this method can provide false positives, it speeds up our heuristics allowing us to run more iterations during the same time limit and improving the solutions found further. We believe that this method could also be interesting in the development of heuristics for other packing problems.

We compare our proposed algorithms with the genetic hybrid algorithm~Hybrid-GA proposed by~\citet{kumar2006scheduling}. We also create an extension of Hybrid-GA for MAXSPACE-RDWV and compare it with our meta-heuristics for this problem. Some meta-heuristics, such as VNS and GRASP$\plus$VNS, have better results than Hybrid-GA for both problems.

In Section~\ref{sec:lr} we present the review of literature for MAXSPACE, MINSPACE and related problems. In Section~\ref{sec:heu} we present our heuristics, in Section~\ref{sec:exp} we present our experiments, and we analyze our results in Section~\ref{sec:res}. 

%% file: revisao.tex
\label{sec:lr}

In this section, we present an overview of the literature for the advertisement scheduling problems. We also provide a brief review of related problems.

We say that an algorithm~$H$ is an~$\alpha$-approximation for a maximization problem if for any instance~$I$ it runs in polynomial time and produces a solution~$S$ such that~${f(S) \ge \alpha\cdot \OPT}$, with~${\alpha \leq 1}$, where~$\OPT$ is the value of the optimal solution of~$I$ and~${f(S)}$ is the value of solution~$S$. For a minimization problem, an~$\alpha$-approximation~$H$ is such that~$H$ is a polynomial algorithm and~${f(S) \leq \alpha\cdot \OPT}$, for~${\alpha \geq 1}$.
A family of algorithms~$\{H_\varepsilon\}$ is a \textit{Polynomial-Time Approximation Scheme}~(PTAS) for a maximization problem if, for every constant~${\varepsilon > 0}$,~$H_\varepsilon$ is a~$(1-\varepsilon)$-approximation. A~\textit{Fully Polynomial-Time Approximation Scheme}~(FPTAS) is a PTAS whose running time is also polynomial in~$1/\varepsilon$~\citep{vazirani2013approximation}. 

Note that MAXSPACE does not admit an FPTAS even for~${K = 2}$, since it generalizes the \textit{Multiple Subset Sum Problem} with identical capacities~(MSSP-I), which does not admit an FPTAS even for~${K = 2}$~\citep{kellerer2004introduction}.

\citet{dawande2003performance} define three special cases of MAXSPACE:~MAX$_w$,~MAX$_{K\mid{w}}$~and~MAX$_s$. In~MAX$_w$, every ad has the same frequency~$w$. In~MAX$_{K\mid{w}}$, every ad has the same frequency~$w$, and the number of slots~$K$ is a multiple of~$w$. And, in MAX$_s$, every ad has the same size $s$. Analogously, they define three special cases of MINSPACE:~MIN$_w$,~MIN$_{K\mid{w}}$~and~MIN$_s$. 

\citet{adler2002scheduling} present a~$\frac{1}{2}$-approximation algorithm called~SUBSET-LSLF for MAXSPACE when the ad sizes form a sequence~${s_1 > s_2 > s_3 > \dots}$, such that for all~$i$,~$s_i$ is a multiple of~$s_{i+1}$. \citet{dawande2003performance} present three approximation algorithms, a~${(\frac{1}{4} + \frac{1}{4K})}$-approximation for MAXSPACE, a~$\frac{1}{3}$-approximation for~MAX$_w$ and a~$\frac{1}{2}$-approximation for~MAX$_{K\mid{w}}$. \citet{freund2002approximating} proposed a~${(\frac{1}{3} - \varepsilon)}$-approximation for MAXSPACE and a~${(\frac{1}{2} - \varepsilon)}$-approximation for the special case in which the size of ads are in the interval~${[L/2, L]}$.

\citet{kumar2006scheduling} present a heuristic for MAXSPACE called \emph{Largest-Size Most-Full}~(LSMF), and use LSMF combined with a genetic algorithm to create a hybrid genetic algorithm to MAXSPACE\@. In the computational experiments, we compare our algorithms with the algorithm proposed by~\citet{kumar2006scheduling}. \citet{amiri2006scheduling} present an integer linear programming for a MAXSPACE variant in which each ad has a set of values for frequency. \citet{da2019polynomial} present a polynomial-time approximation scheme for MAXSPACE with deadlines and release dates and with a constant number of slots. 

Regarding MINSPACE, \citet{adler2002scheduling} present a~$2$-approximation called~\emph{Largest-Size Least-Full} (LSLF) for MINSPACE\@. Algo\-rithm LSLF is also a ${(\frac{4}{3}-\frac{1}{3K/w})}$-approximation to MIN$_{K\mid{w}}$~\citep{dawande2003performance}. \citet{dawande2003performance} present a~$2$-approxi\-mation for MINSPACE using \emph{LP Rounding}, and \citet{dean2003improved} present a~$\frac{4}{3}$-approximation for MINSPACE using Graham's algorithm for schedule~\citep{graham1979optimization}.

Some problems related to MAXSPACE and MINSPACE are the Bin Packing Problem, the Cutting Stock Problem, the Knapsack Problem, and the Multiple Knapsack Problem. The \textit{Bin Packing Problem} (BPP) is defined as follows: given a set of items~${I = \{i_1, i_2 \dots i_n\}}$, where each item~$i$ has height~$h_i$, and an unlimited set of identical bins of height~$H$; allocate all items in bins, in order to minimize the number of bins used, without intersections between items and between items and the border of the bins~\citep{princeton1971performance}.

Some classic algorithms for BPP are \textit{First-Fit} (FF), \textit{Next-Fit} (NF) and \textit{First-Fit Decreasing} (FFD). The~FF algorithm tries to add each item to the first bin that fits, using a new bin when the item does not fit into the bins already used. The~NF algorithm considers putting each item~$i$ in the last bin~$j$ opened. If it is possible to add the item without overflowing the bin capacity,~$i$ is added in~$j$, and the next item is considered, otherwise a new bin~${j+1}$ is created to add~$i$. The~FFD algorithm adds items to containers in non-increasing order of height using \textit{First-Fit}. These heuristics are, respectively, a~$17/10$-approximation, a~$2$-approximation, and a~$11/9$-approximation for BPP~\citep{johnson1973near, johnson1974fast, coffman1984approximation}. 

In the \textit{Cutting Stock Problem}~(CSP), we are given~$m$ items, each having an integer weight~$h_i$ and an integer demand~$d_i$, and an unlimited number of identical bins of integer capacity~$H$. The objective is to pack~$d_i$ copies of each item~$i$ using the minimum number of bins so that the total weight packed in any bin does not exceed its capacity~\citep{delorme2016bin}. In this paper, we use instances from the literature of CSP to compare our algorithms. 

The CSP and the BPP are basically the same problems. They differ essentially in the input items' demand. \citet{cintra2007note} proved that approximation algorithms that the authors define as ``well-behaved'' for BPP could be translated to algorithms with the same approximation ratios for CSP, even with two or three dimensions.

The Cutting Stock Problem has many approaches with linear programming~\citep{gilmore1961linear, dyckhoff1981new, goulimis1990optimal, delorme2016bin} and heuristics~\citep{wascher1996heuristics, lai1997developing, poldi2009heuristics, chen2019genetic}.

The \textit{Knapsack Problem} (KP) consists of, given a container of capacity~$W$ and a set of items~${I = \{i_1, i_2 \dots i_n\}}$, where each item~$i$ has value~$p_i$ and weight~$q_i$, find a subset~${I' \subseteq I}$ of maximum value that does not exceed the capacity of the knapsack, i.e,~${ \sum_{i \in I'}{q_i} \leq W}$~\citep{kellerer2004introduction}.

\citet{ibarra1975fast} and \citet{lawler1979fast} proposed a FPTAS for the KP\@. The Knapsack Problem also has approaches using dynamic programming~\citep{horowitz1974computing, ahrens1975merging, toth1980dynamic} and \textit{branch and bound}~\citep{kolesar1967branch, greenberg1970branch, horowitz1974computing, barr1975linked, nauss1976efficient, martello1977upper, zoltners1978direct}.

The \textit{Multiple Knapsack Problem} (MKP) is a generalization of~KP\@. Given a set of items~${I = \{i_1, i_2 \dots , i_k\}}$, where each item~$i$ has value~$p_i$ and weight~$q_i$, and a set of containers~${M = \{M_1, M_2, \dots , M_m\}}$, where each container~$j$ has a capacity~$c_j$. The MKP consists of finding a subset of items with maximum value with feasible packaging in the containers. If we consider that each container has the same capacity and that each item~$i$ has~$p_i = q_i$, we have the same problem as the special case of MAXSPACE in which each item has only a copy.

\citet{khuri1994zero} presented a genetic algorithm for the MKP, and \citet{kellerer1999polynomial} presented a PTAS for the particular case where all containers have the same capa\-city. \citet{chekuri2005polynomial} showed that MKP does not admit FPTAS even with~$2$ containers and presented a PTAS for the problem.

The MAXSPACE and MINSPACE problems are also related to the \textit{Scheduling Problem}, which consists of, given a set~${S = \{s_1, s_2 \dots s_k\}}$ of tasks, a value~$q_i$ indicating the time needed for the task~$i$ to be entirely executed and a number~$m$ of processors, assign tasks to the processors to minimize the total execution time~\citep{ullman1975np}.

If we consider that each processor is a slot, the sum of the tasks assigned to the processor~$i$ represents the height of the slot~$i$, and that we want to minimize the fullness of the largest slot, we have the same problem as MINSPACE with a single copy per item. If we add a common deadline to all tasks, the goal becomes to maximize the number of tasks that can be entirely executed before the deadline. And so we have a problem similar to MAXSPACE with only a copy per item.

\citet{hochbaum1987using} presented a PTAS for the Scheduling Problem and also showed that this algorithm can be applied in practice to~${\varepsilon = (1/5 + 2^{-m})}$ and~${\varepsilon = (1/6 + 2^{-m})}$, where~$m$ is the number of processors.

%% file: heuristicas.tex
\label{sec:heu}

In this section, we present meta-heuristics for MAXSPACE and its variant, MAXSPACE-RDWV\@. 
We introduce the neighborhood structures used and the procedure for constructing initial solutions. Then, we present the Tabu Search, VNS, and GRASP metaheuristics. We developed three GRASP versions: using Tabu Search, VNS, and best improvement as local search procedure. Tabu Search was used only as a GRASP subroutine since it did not perform well independently in preliminary experiments, while VNS was also executed separately from GRASP\@.

\subsection{Neighborhoods\label{sec:neigh}}

The meta-heuristics applied in this work are based on local search. Next, we present the neighborhood structures used by our local search-based algorithms. In these neighborhood structures, we only consider feasible moves.

\textbf{ADD$(A_i)$}: add an unscheduled advertisement~$A_i$ to the current solution. Each movement in this neighborhood corresponds to an ad~$A_i$ that can be added to the solution, that is, it is possible to add at least~$w^{min}_{i}$ copies of~$A_i$ to the current solution and keep it feasible. The ad is placed by a first-fit heuristic, which adds a copy to the first slot with no copies of~$A_i$ without exceeding the capacity of the slot while not violating the restrictions of release date and deadline. This neighborhood tries to insert as many copies of~$A_i$ as possible, without exceeding~$w^{max}_{i}$. 

\textbf{CHG${(A_i, A_j)}$}: remove an ad~$A_i$ scheduled in the current solution and add an advertisement~$A_j$ that is not scheduled. In this structure, we consider only valid changes, that is: when it is possible to add~$A_j$ in the solution after removing~$A_i$. We add~$A_j$ using~ADD neighborhood. Notice that generating all the neighbors of this structure to a solution is very expensive.

\textbf{RPCK${(A_i^l, A_j^u)}$}: changes the~$l$-th copy of an ad~$A_i$ in the solution to the~$u$-th copy of an ad~$A_j$ that is also in the solution. The goal with this neighborhood is to repack the copies of ads in some solution to open space to other ads or copies that can be added.

\textbf{ADDCPY${(A_i, l, j)}$}: add the~$l$-th copy of an ad~$A_i$, which is in the solution to slot~$j$. This neighborhood is only applied to MAXSPACE-RDWV in which the ad has a frequency between~$w^{min}_{i}$ and~$w^{max}_{i}$. The idea is to try to add one more copy of an ad that has at least~$w^{min}_{i}$ copies in the current solution, but does not have~$w^{max}_{i}$ copies scheduled yet. 

\textbf{MV${(A_i^l, j)}$}: move the~$l$-th copy of an ad~$A_i$ in the solution to slot~$j$. As in~RPCK neighborhood, we want to repack the copies of ads in some solution to open space to other ads or copies that can be added. 


\subsubsection{Secondary Objective Function}

We note that RPCK and MV do not change the value of the solution. Thus, in order to guide the search on these neighborhoods, we use a secondary objective function the sum of the square of empty spaces in the slots, that is, let~$f(B_j)$ be the fullness of a slot~$j$ we use
\begin{equation}
\label{eq:obj}
    \sum^{K}_{j=1}{{(L - f(B_j))}^2}
\end{equation}
as the objective function. 

During our algorithms, we alternate between minimizing and maximizing this objective function. When minimizing the objective, we try to level the fullness of the slots; that is, we avoid full slots when the ads could be distributed to the empty ones. With that, we try to open space to add a new ad to the solution. And when we maximize this objective, we want to fill some slots as much as possible, even if others are empty, trying to add an advertisement with few copies of considerable size or add more copies of scheduled ads in MAXSPACE-RDWV\@.

\begin{figure}[H]
    \centering
    \includegraphics[width=\textwidth]{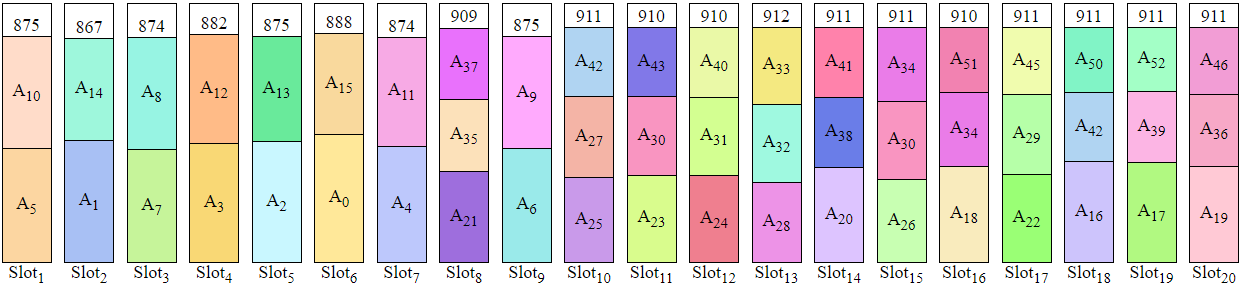}
    \caption{Solution for instance~\emph{Falkenauer\_t60\_12} obtained by GRASP+VNS minimizing function~\eqref{eq:obj}. This solution has value~$17938$.}
    \label{fig:phase1}
\end{figure}

\begin{figure}[H]
    \centering
    \includegraphics[width=\textwidth]{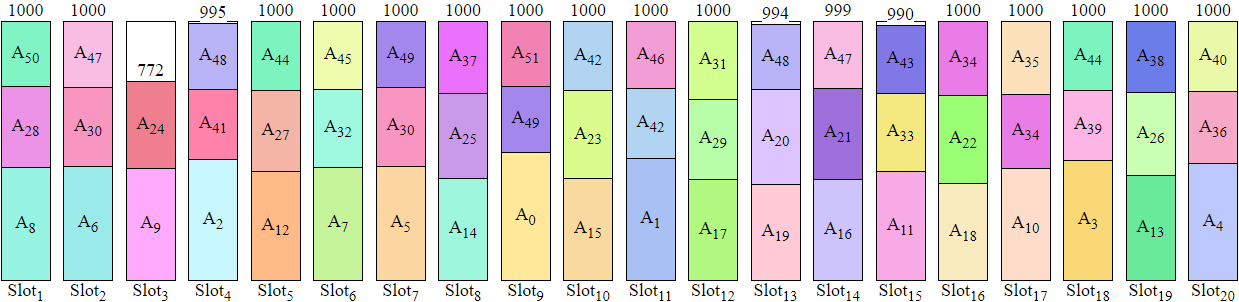}
    \caption{Solution for instance~\emph{Falkenauer\_t60\_12} obtained by GRASP+VNS maximizing function~\eqref{eq:obj}. This solution has value~$19750$.}
    \label{fig:phase2}
\end{figure}

\begin{figure}[H]
    \centering
    \includegraphics[width=\textwidth]{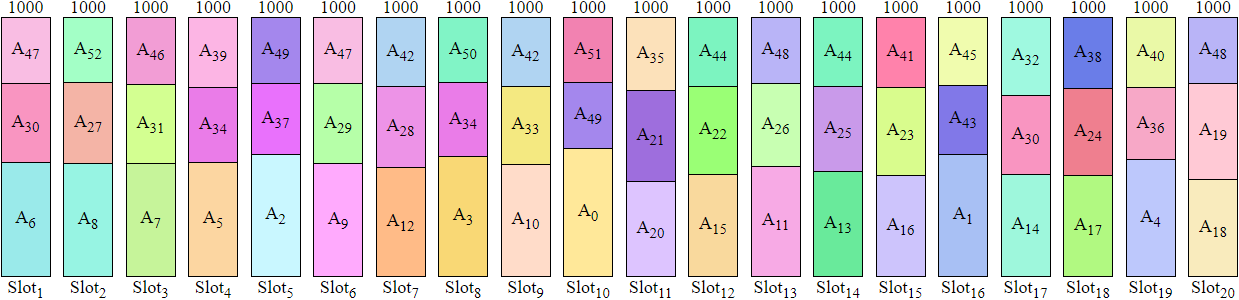}
    \caption{Solution for instance~\emph{Falkenauer\_t60\_12} obtained by GRASP+VNS alternating between minimizing and maximizing function~\eqref{eq:obj}. This is an optimal solution, with value~$20000$.}\label{fig:dualphases}
\end{figure}

Figures~\ref{fig:phase1},~\ref{fig:phase2} and~~\ref{fig:dualphases} show solutions for instance~\emph{Falkenauer\_t60\_12} obtained by GRASP$\plus$VNS minimizing function~\eqref{eq:obj}, maximizing function~\eqref{eq:obj} and alternating between minimizing and maximizing function~\eqref{eq:obj}, respectively. Note that in the solution of Figure~\ref{fig:phase1}, the slots have level fullness, and in the solution of Figure~\ref{fig:phase2}, most of the slots are full, but there is a slot with little fullness. In the solution in Figure~\ref{fig:dualphases}, we merge the ideas used in the previous solutions to obtain an optimal solution for the instance.

\subsubsection{Feasibility Check for ADD}

In implementing the ADD neighborhood for MAXSPACE-RDWV, we use a Binary Indexed Tree~(BIT) introduced by~\citet{fenwick1994new} to do some verifications more efficiently. This tree allows us to verify the sum of an interval in a vector of~$d$ integers in time complexity~${\cal{O}}(\log d)$. It also allows us to verify the maximum or minimum value in an interval in time complexity~${\cal{O}}(\log d)$, as shown in~\citet{dima2015efficient}. The time complexity to create a BIT is~${\cal{O}}(d)$, and to update it is~${\cal{O}}(\log d)$.

We use a BIT to verify the space remaining in an interval of slots. If we want to add~$t$ copies of an ad~$A_i$, we can ascertain in~${\cal{O}}(\log K)$ using the BIT if the space remaining in the interval of slots~${[r_i, d_i]}$ is at least~$t\,s_i$. We can also verify in~${\cal{O}}(\log K)$ if the least full slot in this interval has enough space to store a copy of~$A_i$. 

This method can provide false positives, that is, it can be impossible to pack $A_i$ even if the remaining space is at least $t\,s_i$ and the least full slot has a remaining space of at least $s_i$. Nonetheless, it cannot provide false negatives and, thus, it is used to speed up the ADD neighborhood.

\begin{figure}[H]
    \centering
    \includegraphics[width=0.6\textwidth]{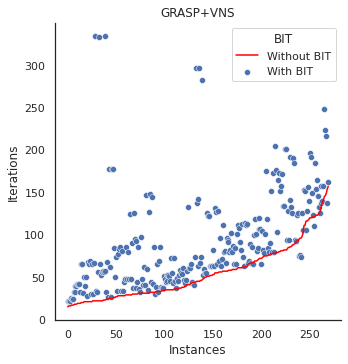}
    \caption{Comparison between GRAPS+VNS with BIT and GRAPS+VNS without BIT}
    \label{fig:BIT}
\end{figure}

In Figure~\ref{fig:BIT}, we compare the number of iterations of GRASP$\plus$VNS using BIT with GRASP$\plus$VNS without BIT. This graph considers only instances where GRASP$\plus$VNS without BIT executes at least~$10$ iterations. The red line represents the number of iterations GRASP$\plus$VNS without BIT executes for each instance, and the blue points represent the number of iterations GRASP$\plus$VNS with BIT executes in the same instances. The instances were sorted by the number of iterations executed by GRASP$\plus$VNS without BIT. When a blue point is over the red line, it means that in that instance, GRASP$\plus$VNS with BIT executed more iterations than GRASP$\plus$VNS without BIT. Observe that GRAPS$\plus$VNS with BIT executed more iterations in almost all instances.

\subsection{Constructive Heuristic}\label{sec:hc}

In this section, we present the heuristic used to construct initial solutions for our algorithms. 

The constructive heuristic takes a parameter~$\alpha$ that has a value in the range~$[0, 1]$ that indicates how greedy or random it will be. The closer to~$0$ the value of~$\alpha$ is, the greedier the construction heuristic is, and the closer to~$1$, the more random it is.

At each iteration, the subroutine selects a set~$C$ of candidates and calculates the cost of each in relation to the solution under construction~$S$. Candidate ads are those not in the solution under construction~$S$, which can be added to~$S$. 
From the list of candidates and the calculated costs, the algorithm creates a restricted list of candidates~$RC$ with only the best candidates. That is, let~$max$ and~$min$ be the maximum and minimum cost of ads not scheduled yet and, given an~${\alpha \in [0,1]}$. The algorithm randomly chooses an ad~$A_j$ from the set of ads with cost in interval~\mbox{$[max-\alpha(max-min), max]$}. After creating the shortlist of candidates, an ad~$A_j$ of~$RC$ is chosen uniformly at random to be part of~$S$, and the procedure is repeated. The chosen ad~$A_j$ is added using first fit if possible, otherwise~$A_j$ is discarded. The algorithm ends when there are no more candidates, and the solution~$S$ is returned. We consider the cost of each ad~$A_i$ as~$s_i w_i$ in MAXSPACE and~$v_i/s_i$ in MAXSPACE-RDWV\@. Algorithm~\ref{alg:hc} presents the pseudocode of this constructive heuristic.

\begin{scriptsize}
\begin{algorithm}[H]
\caption{Constructive Heuristic\label{alg:hc}}
\begin{algorithmic}[1]

\Procedure{ConstructiveHeuristic}{$\alpha$}
    \State $S \gets \emptyset$
    \State $C \gets \cala$
    \For{$A_i \in C$}
        \State $cost(A_i) \gets s_i w_i$ \Comment{in MAXSPACE-RDWV the cost is $v_i/s_i$}
    \EndFor
    \While{$C \neq \emptyset$}
        \State $min \gets \min_{A_i \in C}\{cost(A_i)\}$
        \State $max \gets \max_{A_i \in C}\{cost(A_i)\}$
        \State $RC \gets \emptyset$
        \For{$A_i \in C$}
            \If{$cost(A_i) \geq max-\alpha(max-min)$}
                \State $RC \gets RC \cup \{A_i\}$
            \EndIf
        \EndFor
        \State Randomly choose $A_j$ from $RC$
        \State $S \gets \Call{FirstFit}{A_j, S}$ \Comment{Only if~$A_j$ fits}
        \State $C \gets C \setminus \{A_j\}$
    \EndWhile
    \State \Return $S$
\EndProcedure

\end{algorithmic}
\end{algorithm}
\end{scriptsize}

\subsection{Tabu Search}\label{sec:ts}

Tabu Search~(TS) is a meta-heuristic proposed by~\citet{glover1986future} that allows Local Search to overcome local minima. This meta-heuristic memorizes the improvements of Local Search in a structure called Tabu List and forbid moves while they are on this list. Each element is kept in the Tabu List until some quantity of improvements is reached~\citep{gendreau2010handbook}.

The construction of the initial solution for Tabu Search was made using the constructive heuristic present in Section~\ref{sec:hc}. The Tabu Search uses all neighborhood structures present in Section~\ref{sec:neigh} and has a maximum number of iterations as a stopping condition. Three versions of Local Search were designed. The first one randomly chooses a neighborhood at each iteration. The second one starts in neighborhood 1 and only changes when no improvement is found. The third version circularly chooses the next neighborhood.

The Tabu List memorizes only previous moves since it would be expensive to store the complete solutions, i.e., for the neighbor~CHG${(A_i, A_j)}$ the list only memorizes the type of this neighbor and the ads~$A_i$ and~$A_j$, and it keeps this movement forbid while it is in the list.

Our Tabu Search has two phases: the first one minimizes the objective function in the RPCK and MV neighborhood structures, and the second one maximizes these objectives. The algorithm switches between the first and second phases while an improvement is found.

\subsection{VNS}\label{sec:vns}

The Variable Neighborhood Search~(VNS) is a meta-heuristic proposed by~\citet{mladenovic1997variable} to solve optimization problems. It consists of a descent phase with systematic changes of the neighborhood to find local minima, and a perturbation phase (shaking) to escape valleys. A neighborhood structure for a given optimization problem can be defined as~${\caln_k}$ (for~${k=1, \dots, k_{max}}$) and~$\caln_k(S)$ denotes the set of solutions in~$k$-th neighborhood of a solution~$S$~\citep{gendreau2010handbook}.

For MAXSPACE, the neighborhood structures were considered in the following order: MV, RPCK, ADD, and CHG, and for MAXSPACE-RDWV, the neighborhood structures were considered in the following order: MV, RPCK, ADDCPY, ADD and CHG\@. The order of the neighborhoods was defined considering the cost of calculating them, leaving the most costly neighborhoods to the end, which makes them be explored less often.

The construction of the initial solution for VNS was made using the constructive heuristic present in Section~\ref{sec:hc}. In the local search step, our VNS uses a Variable Neighborhood Descent~(VND) meta-heuristic, which is a version of VNS in which the change of neighborhoods is performed in a deterministic way~\citep{gendreau2010handbook}. Also, VNS shaking has been changed to perform~$\mathcal{Q}$ disturbances before switching neighborhoods, to increase the chance to escape local minima.

Our VNS also has two phases, minimizing and maximizing the objective function in the RPCK and MV neighborhood structures. The algorithm switches between the first and second phases while an improvement is found.

\subsection{GRASP}

The Greedy Randomized Adaptive Search Procedure~(GRASP) is a meta-heuristic that creates good random initial solutions and uses a local search to improve them~\citep{feo1995greedy}. The GRASP executes~$k$ iterations, and at each iteration, produces a random initial solution and runs a local search to improve it. The algorithm returns the best solution found in~$k$ iterations.

We use the constructive heuristic present in Section~\ref{sec:hc} to create initial solutions at each iteration of GRASP\@. Three versions of the GRASP were designed. The first one uses a local search with the best improvement, the second one uses VNS (as present in Section~\ref{sec:vns}) as a local search procedure, and the third version uses Tabu Search (as present in Section~\ref{sec:ts}) as local search procedure.

The local search methods used in our GRASPs have two phases, as shown in Sections~\ref{sec:ts} and~\ref{sec:vns}.

%% file: experimentos.tex
\label{sec:exp}
This section presents the instances used, the selected parameters, and how the tests were performed.

\subsection{Instances}

Instances were randomly generated with uniform probability and were divided into~$36$ sets. These sets of instances are related to size, frequencies, profits, release dates, and deadline of ads. Three types of ads size were considered: small, medium, and large. An ad is called small if it has~$s_i$ in interval~${[1, L/4]}$, is called medium if it has~$s_i$ in interval~${(L/4, L/2]}$, and is called large if it has~$s_i$ greater than~$L/2$. We also consider three frequencies for ads: infrequent, medium frequency, and very frequent. An ad is called infrequent if it has~$w^{min}_{i}$ in interval~${[1, 5]}$ and has~$w^{max}_{i}$ in interval~${[6, 10]}$, is called medium frequency if it has~$w^{min}_{i}$ in interval~$[11, 15]$ and has~$w^{max}_{i}$ in interval~${[16, 20]}$, and is called very frequent if it has~$w^{min}_{i}$ in interval~${[21, 25]}$ and has~$w^{max}_{i}$ in interval~${[26, 30]}$. These values for frequencies were chosen based on instances of~\citet{amiri2006scheduling}. We consider two values for profits of ads: related to the size (with~${v_i = s_i}$) and random (with~$v_i$ in the interval~${[1, 100]}$). Moreover, we also consider instances without release dates and deadlines and with release dates and deadlines randomly chosen (choosing~$r_i$ from interval~${[1, K - w_i^{min}]}$ and~$d_i$ from interval~${[r_i + w_i^{min}, K]}$).

Instances of~$4$ different sizes were generated, according to Table~\ref{tamanhos}, for each set were generated~$10$ instances of each size, which give us~$40$ instances per set.

\renewcommand{\|}{\text{\textbar}}
\begin{table}[ht]
    \caption{Size of generated instances.\label{tamanhos}}
    \centering
    \begin{tabular*}{\textwidth}{c @{\extracolsep{\fill}}cc}
        \toprule
        \textbf{$\|\cala\|$} & \textbf{$K$} & \textbf{$L$} \\ \midrule
        $100$              & $75$         & $50$         \\
        $500$              & $250$        & $100$        \\
        $1000$             & $500$        & $250$        \\
        $10000$            & $500$        & $200$        \\ \bottomrule
    \end{tabular*}
\end{table}

In addition to randomly generated instances, we use all instances provided by the \textit{Bin Packing Problem Library} benchmark~\citep{delorme2018bpplib} for the Cutting Stock Problem~(CSP). The size of the slots was set to be the same as the capacity of the containers in the CSP, i.e.,~${L = H}$. The number of slots~$K$ was defined as~$\sum_{i \in I}{d_i}/3$ in the Falkenauer Triples class and~${\lceil \sum_{i \in I}{h_i d_i}/L \rceil}$ in the other classes. Each item~$i$ in an original~CSP instance was mapped to an ad~$A_i$ in the generated instance as follows:~${w_i = \min\{d_i, N\}}$ and~${s_i = h_i}$. We use literature instances only for MAXSPACE\@. The MAXSPACE-RDWV experiments were performed only with random instances because the insertion of release dates, deadlines, variable frequency, and value can make the instances lose their combinatorial structure. Without the insertion of such attributes, the problem is identical to MAXSPACE\@.

In Table~\ref{instances}, we present the number of instances for each problem in each class.

\begin{table}[ht]
\caption{Number of instances in each class.\label{instances}}
\centering
\begin{tabular*}{\textwidth}{l @{\extracolsep{\fill}}cccc}
 \toprule                                                                                           & \multicolumn{2}{c}{\textbf{MAXSPACE}} & \multicolumn{2}{c}{\textbf{MAXSPACE-RDWV}} \\ 
\multicolumn{1}{l}{\textbf{Instance Class}}                                                         & \textbf{\#}          & \%              & \textbf{\#}          & \textbf{\%}          \\ 
\midrule
\multicolumn{1}{l}{Random}                                                               & 1080                 & 31.44           & 360                  & 100.00                \\ 
\multicolumn{1}{l}{\citet{delorme2016bin}}                                                              & 500                  & 14.56           & 0                    & 0                \\ 
\multicolumn{1}{l}{Falkenauer Triples~\citep{falkenauer1996hybrid}}                                                   & 80                   & 2.33            & 0                    & 0                 \\ 
\multicolumn{1}{l}{Falkenauer Uniforms~\citep{falkenauer1996hybrid}}                                                  & 80                   & 2.33            & 0                    & 0                 \\ 
\multicolumn{1}{l}{\citet{schoenfield2002fast}}                                                              & 28                   & 0.82            & 0                    & 0                 \\ 
\multicolumn{1}{l}{\citet{gschwind2016dual}}                                                               & 240                  & 6.99            & 0                    & 0                 \\ 
\multicolumn{1}{l}{Scholl~1 and 2~\citep{scholl1997bison}}                                            & 1200                 & 34.93           & 0                    & 0                \\ 
\multicolumn{1}{l}{Scholl~3~\citep{scholl1997bison}}                                              & 10                   & 0.29            & 0                    & 0                 \\ 
\multicolumn{1}{l}{\citet{schwerin1997bin}}                                                             & 200                  & 5.82            & 0                    & 0                 \\ 
\multicolumn{1}{l}{\citet{wascher1996heuristics}}                                                             & 17                   & 0.49            & 0                    & 0                 \\ 
\multicolumn{1}{l}{\textbf{Total}}                                                       & 3435                 & 100.00          & 360                  & 100.00               \\ \bottomrule
\end{tabular*}
\end{table}

\subsection{Choosing Parameters and Running Experiments}

Algorithm Hybrid-GA~\citep{kumar2006scheduling} was implemented to be compared with our heuristics. This algorithm was initially proposed for MAXSPACE, but we also developed a version of it for MAXSPACE-RDWV, adding the restrictions of release dates and deadlines and adding copies of an ad~$A_i$ while it is possible (without exceeding~$w_i^{max}$ copies).

Before running the experiments, we used Irace~$2.0$~\citep{lopez2016irace} to choose the parameters of the algorithms. The interval considered for~$\alpha$ was~${[0, 1]}$, for~$\mathcal{Q}$ was~${[1, 10]}$, for~$|tabuList|$ was~${[5, 100]}$ and for the number of Tabu search iterations was~${[50, 500]}$. The precision considered for decimal values was one decimal place.

We gave a timeout of~$5$ days for Irace to select the parameters of each algorithm. Tables~\ref{parametros_max} and~\ref{parametros_maxrdwv} show the chosen parameters for, respectively, MAXSPACE and MAXSPACE-RDWV\@. The number of GRASP iterations was chosen from the preliminary executions of the algorithms. The mean of the iterations in which the best solution was found plus three times the standard deviation was used.

\begin{table}[ht]
    \caption{Parameters chosen by Irace for MAXSPACE.\label{parametros_max}}
    
    \centering
    \begin{tabular*}{\textwidth}{c @{\extracolsep{\fill}} cccccc}
        \toprule
        \textbf{Algorithm} & \textbf{$\alpha$} & \textbf{\# iterations} & \textbf{$\mathcal{Q}$} & \textbf{$|tabu|$} & \textbf{\# iterations of TS} & \textbf{TS type}\\ \midrule
        VNS  & $0.2$ & N/A  & $8$  & N/A  & N/A   & N/A  \\
        GRASP  & $0.3$   & $2000$  & N/A  & N/A   & N/A  &  N/A \\
        GRASP$\plus$Tabu  & $0.9$ & $2000$ & N/A & $55$ & $60$ & Version 2 \\
        GRASP$\plus$VNS   & $0.5$  & $1000$  & $10$  & N/A   & N/A    & N/A \\ \bottomrule
    \end{tabular*}
\end{table}

\begin{table}[ht]
    \caption{Parameters chosen by Irace for MAXSPACE-RDWV.\label{parametros_maxrdwv}}
    
    \centering
    \begin{tabular*}{\textwidth}{c @{\extracolsep{\fill}} cccccc}
        \toprule
        \textbf{Algorithm} & \textbf{$\alpha$} & \textbf{\# iterations} & \textbf{$\mathcal{Q}$} & \textbf{$|tabu|$} & \textbf{\# iterations of TS} & \textbf{TS type}\\ \midrule
        VNS & $0$ & N/A & $5$ & N/A & N/A & N/A \\
        GRASP & $0.3$ & $2000$ & N/A & N/A & N/A & N/A \\
        GRASP$\plus$Tabu & $0.2$ & $2000$ & N/A & $100$ & $320$ & Version 3 \\
        GRASP$\plus$VNS  & $0.2$ & $2000$ & $9$ & N/A  & N/A  & N/A \\ \bottomrule
    \end{tabular*}
\end{table}

The parameters used in Hybrid-GA were also obtained from Irace and the timeout was~$5$ days for each version. The interval considered for population size~$p_s$ was~${[100, 1000]}$, for the fraction of elites~$\varepsilon$ the interval considered was~${[0.1, 0.35]}$, for the crossover probability~$p_c$ was~${[0.65, 0.80]}$, for the mutation probability~$p_m$ was considered the interval~${[0.1, 0.25]}$, the number of independent population~$P$ was chosen from interval~${[1, 3]}$ and the number of generations~$n_{gen}$ was chosen from interval~${[100, 300]}$. For MAXSPACE the chosen parameters were:~${p_s = 400}$,~${\varepsilon = 0.2}$,~${p_c = 0.7}$,~${p_m = 0.1}$,~${P = 3}$ and~${n_{gen} = 300}$. And for MAXSPACE-RDWV Irace chooses the parameters:~${p_s = 350}$,~${\varepsilon = 0.2}$,~${p_c = 0.7}$,~${p_m = 0.2}$,~${P=1}$ and~${n_{gen} = 230}$.

The algorithms were implemented in C$\plus \plus$, and the experiments have been performed in a machine with Intel(R) Xeon(R) Silver 4114 CPU @ 2.20 GHz, 32 GB of memory and Linux OS\@. The timeout for each execution was~$600$ seconds. The seed used to generate random numbers was~$0$. We use a constant seed to allow the experiments to be replicated.

%% file: resultados.tex
In this section, we present and discuss the computational results of the implemented heuristics. We present a separate analysis for MAXSPACE and MAXSPACE-RDWV\@. In both, we compared the results with the Hybrid-GA algorithm~\citep{kumar2006scheduling}. We also present statistical analysis to show that there is a statistical difference between our heuristics and Hybrid-GA\@.

\subsection{MAXSPACE}

\begin{figure}[H]
    \centering
    \includegraphics[width=.7\textwidth]{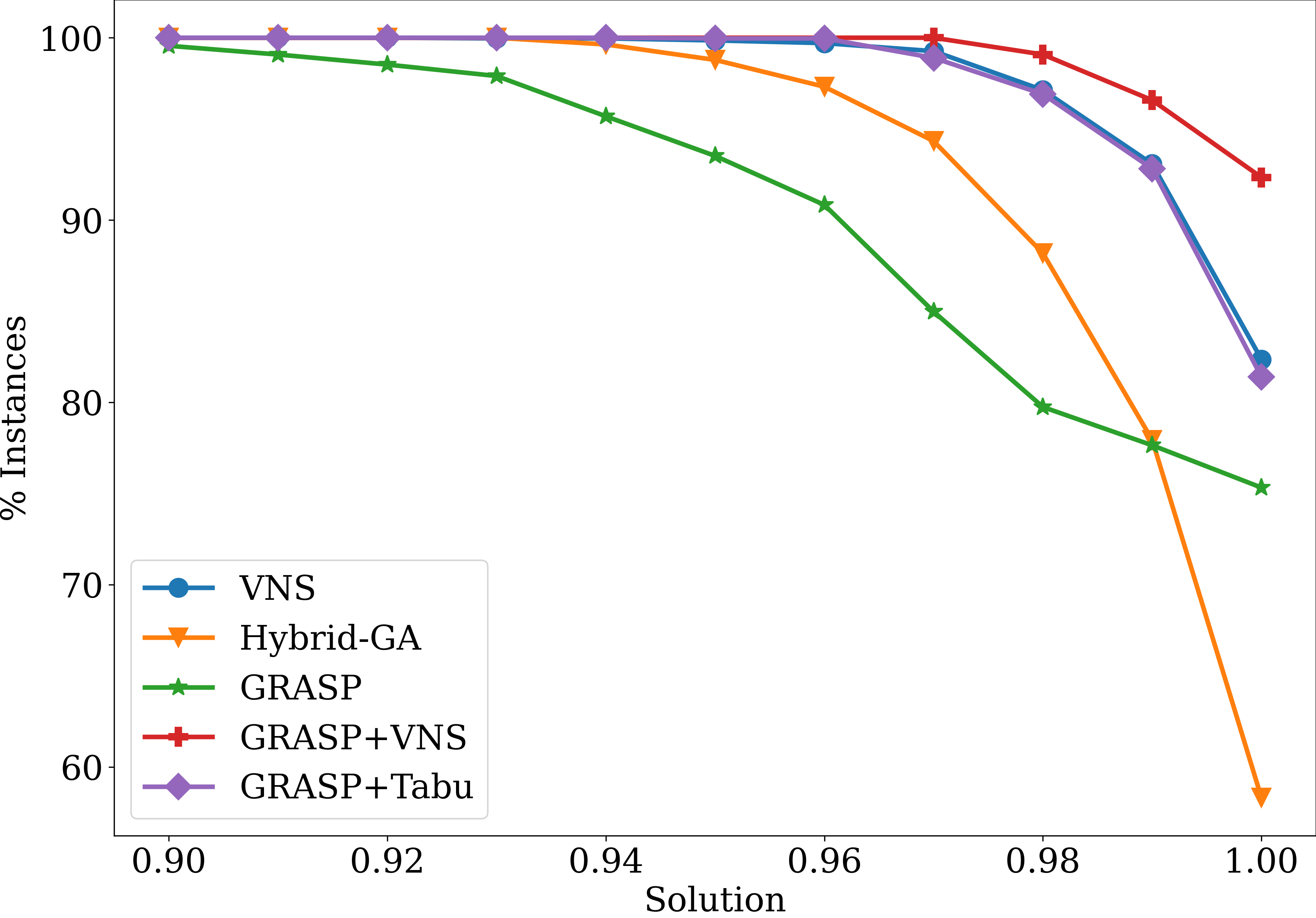}
    \caption{Profiling graph for MAXSPACE.\label{fig:resultados_MAX}}
\end{figure}

First, we analyze the results for MAXSPACE\@. Figure~\ref{fig:resultados_MAX} presents a performance profile~\citep{dolan2002benchmarking} with a comparison of solutions founds by the implemented algorithms for MAXSPACE considering the whole set of instances. The~$x$-axis of the graph represents the quality of the solution relative to the best solution found among all algorithms, and the~$y$-axis represents the percentage of instances the algorithm has achieved such quality. 
For example, if we look at~${x = 0.96}$, the~$y$-axis indicates the percentage of instances each algorithm has reached at least the~$0.96$ of the best solution found by the algorithms. We can observe in the graph that Hybrid-GA reached at least~$0.93$ of the best solution value in the whole set of instances, but reached the best solution only in~$60\%$ of instances. The heuristic GRASP$\plus$VNS achieved a solution quality of at least~$0.97$ for the whole set of instances and found the best solution in more than~$95\%$ of instances, which is the best percentage among the compared algorithms. 

\begin{figure}[H]
    \centering
    \includegraphics[width=0.9\textwidth]{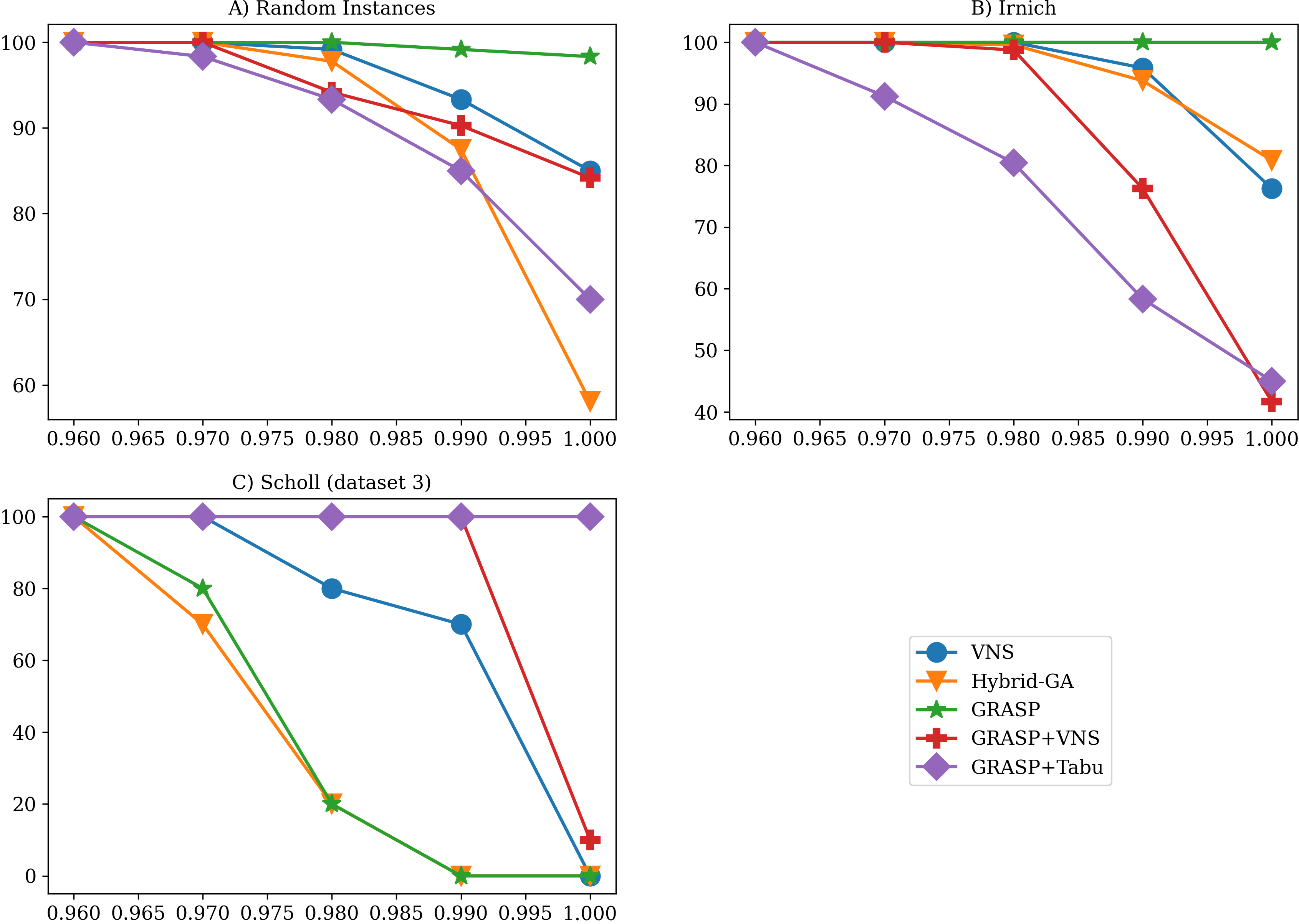}
    \caption{Datasets in which GRASP$+$VNS did not obtain the best results for MAXSPACE\@.\label{fig:MAX_multiplot}}
\end{figure}

In Figure~\ref{fig:MAX_multiplot}, we present profiling graphs considering only datasets of instances in which GRASP$\plus$VNS did not get the best results for MAXSPACE\@. Also, for these datasets, one of our algorithms obtained better solutions than Hybrid-GA (GRASP for random instances and dataset of \citet{gschwind2016dual}, and GRASP$\plus$Tabu for dataset~$3$ of \citet{scholl1997bison}). For the others datasets, the profiling graphs are similar to the general chart present in Figure~\ref{fig:resultados_MAX} and therefore were omitted.

The dataset of~\citet{gschwind2016dual} was generated selecting values randomly in defined intervals for items values, items length, and capacities of bins, similar to the way we developed the random instances in this work. In general, random instances are easier to be solved, which explains how GRASP obtains good results for these instances (Figure~\ref{fig:MAX_multiplot}A and Figure~\ref{fig:MAX_multiplot}B). 
In these instances, the local search process used in the other algorithms may be time-consuming without improving the solution. At the same time, GRASP generates several initial solutions that are already good enough and chooses the best one.

In dataset~$3$ of~\citet{scholl1997bison} the weights are widely spread, and the number of items per bin lies between~$3$ and~$5$. The initial solutions may not be good enough with these characteristics and need a more sophisticated local search process to get more significant improvements. This explains why GRASP obtained results well below the algorithms that use Tabu Search and VNS as the local search process in this set of instances (Figure~\ref{fig:MAX_multiplot}C).

\smallskip

\begin{figure}[H]
    \centering
    \includegraphics[width=.7\textwidth]{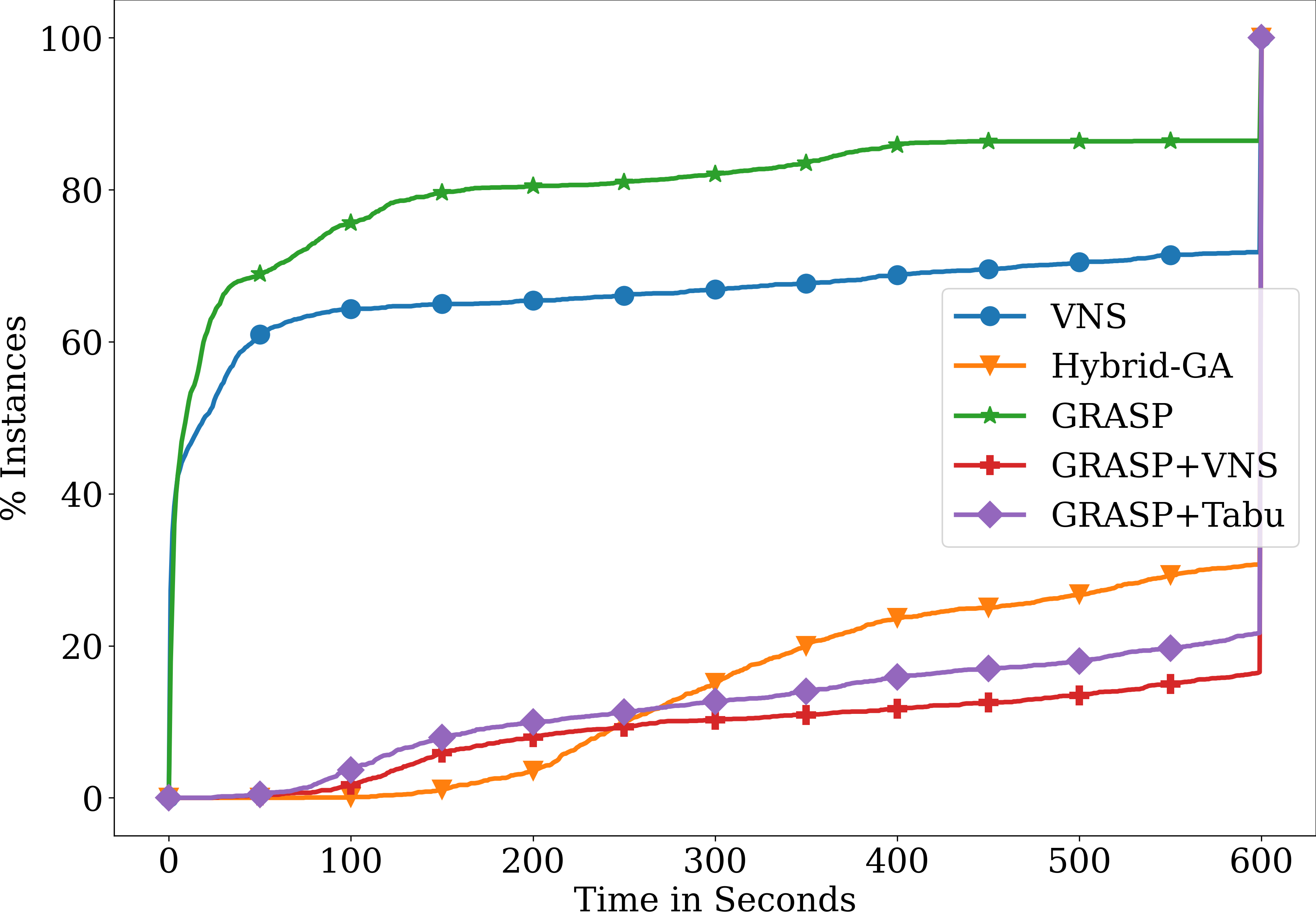}
    \caption{Graph of time for MAXSPACE\@.\label{fig:resultados_MAX_tempo}}
    
\end{figure}

In Figure~\ref{fig:resultados_MAX_tempo}, we present a graph of time of the developed algorithms for MAXSPACE\@. The~$x$-axis of the graph present the time in seconds, and the~$y$-axis show the percentage of instances the algorithm ends within such time. For example, if we observe the~${x = 200}$, the~$y$-axis indicates the percentage of instances in which the algorithm ends in at most~$200$ seconds. We can see in the graph that GRASP ends within~$100$ seconds for~$80\%$ of instances. GRASP$\plus$Tabu, GRASP$\plus$VNS and Hybrid-GA were the most time-consuming algorithms, reaching the timeout of~$600$ seconds in at least~$75\%$ of executions.

Based on the graphs present before, we considered VNS and GRASP$\plus$VNS as our best algorithms for MAXSPACE\@. In Table~\ref{tab:comp_max}, we present a comparison of solutions value among these algorithms and Hybrid-GA for MAXSPACE\@. Each cell of this table presents how many instances the algorithm of a line found a solution better than the solution found by the algorithm of a column. 

\begin{table}[ht]
\centering
\caption{Comparison of algorithms solutions for MAXSPACE.}\label{tab:comp_max}
\begin{tabular*}{\textwidth}{c @{\extracolsep{\fill}}ccc}
\toprule%
& \textbf{Hybrid-GA} & \textbf{VNS} & \textbf{GRASP$\plus$VNS} \\ \midrule

\multicolumn{1}{c}{\textbf{Hybrid-GA}}  & -           & 569            & 450     \\ 
\multicolumn{1}{c}{\textbf{VNS}}      & 2113        & -              & 572     \\ 
\multicolumn{1}{c}{\textbf{GRASP$\plus$VNS}}  & 2244        & 1916           & -        \\ 
\bottomrule
\end{tabular*}
\end{table}

In Table~\ref{tab:comp_max}, we observe that GRASP$\plus$VNS is the algorithm that obtains the best solutions for MAXSPACE, with~$1916$ better solutions than Hybrid-GA and~$2244$ better solutions than VNS\@.

\subsection{MAXSPACE-RDWV}

\begin{figure}[H]
    \centering
    \includegraphics[width=.7\textwidth]{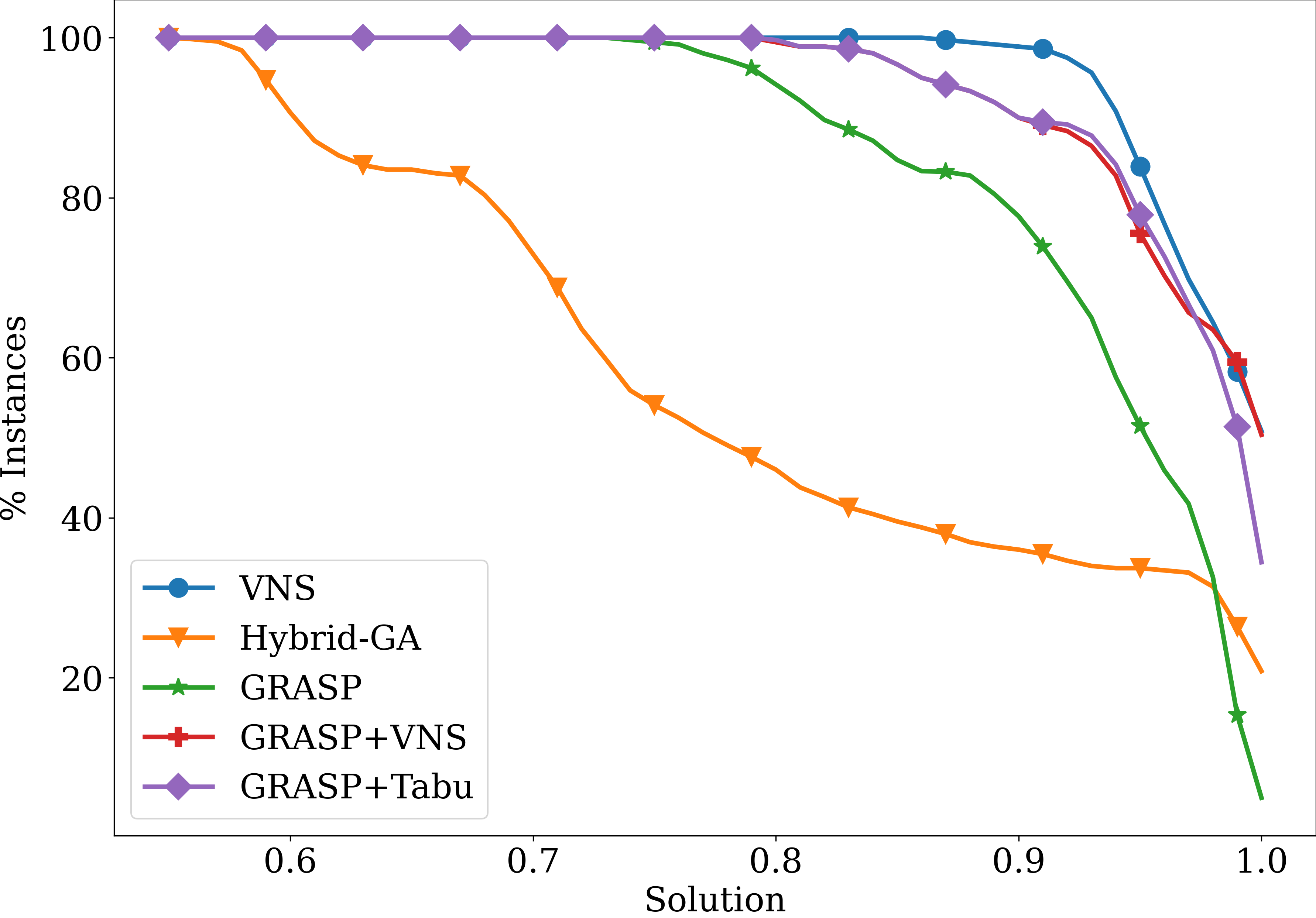}
    \caption{Profiling graph for MAXSPACE-RDWV.\label{fig:resultados_MAX-RDWV-55}}
\end{figure}

\begin{figure}[H]
    \centering
    \includegraphics[width=.7\textwidth]{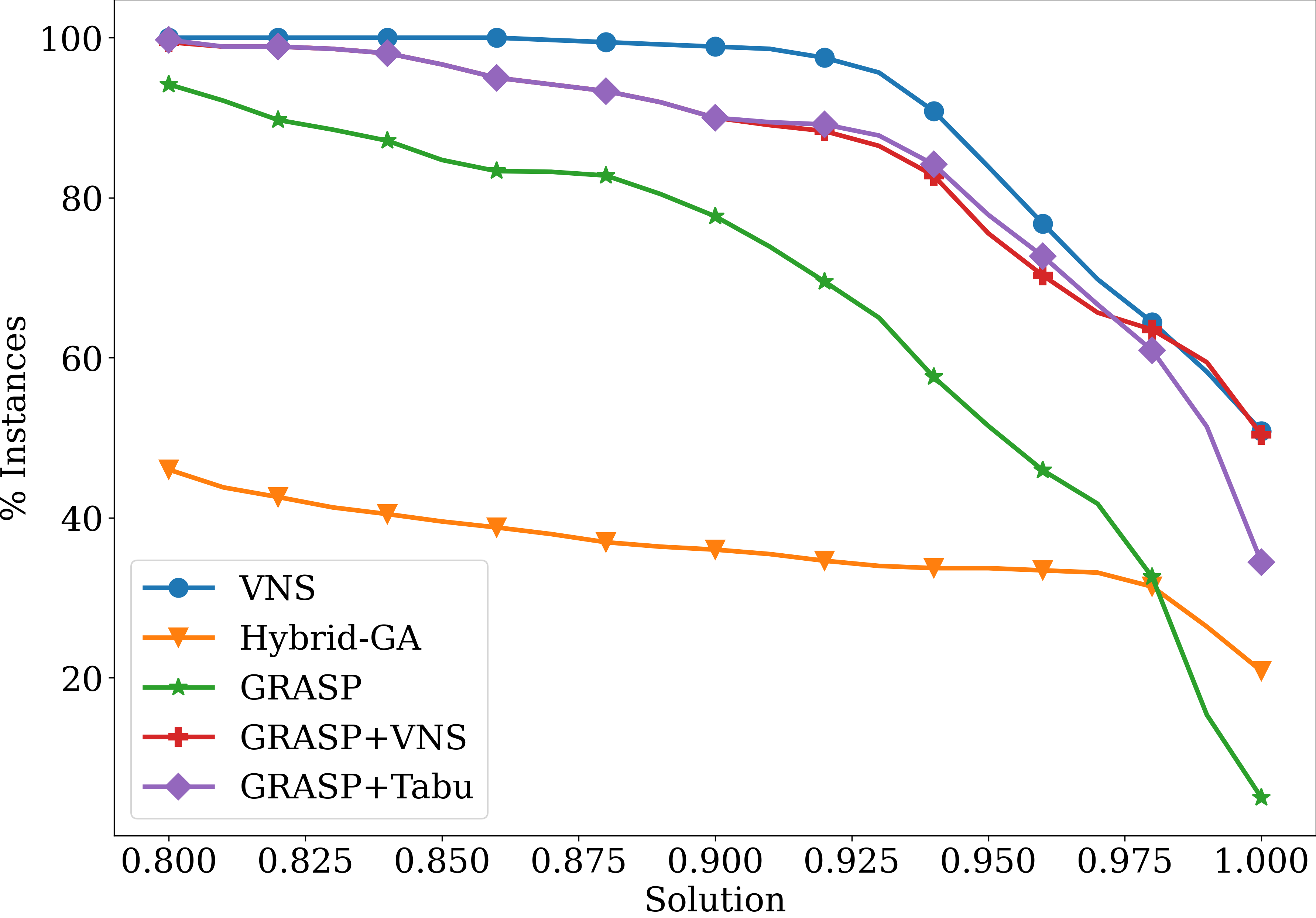}
    \caption{Profiling graph for MAXSPACE-RDWV with~$x$ at least~$0.8$.\label{fig:resultados_MAX-RDWV}}
\end{figure}

In this section, we analyze the results of the heuristics for MAXSPACE-RDWV\@. In Figure~\ref{fig:resultados_MAX-RDWV-55}, we present a profiling graph comparing the solutions found by the algorithms implemented for MAXSPACE-RDWV and in Figure~\ref{fig:resultados_MAX-RDWV} we present a version of this chart with a smaller range of the~$x$ axis for more comfortable viewing.

In the graph of Figure~\ref{fig:resultados_MAX-RDWV-55}, it can be seen that the proposed heuristics obtained better quality than the Hybrid-GA algorithm, which guaranteed the solution quality of only~$0.55$ for the whole set of instances and  achieved the best solution only in approximately~$20\%$ of instances. In the graph in Figure~\ref{fig:resultados_MAX-RDWV}, we can see that VNS guaranteed the best results, with a solution quality of at least~$0.85$ and finding the best solution in approximately~$50\%$ of instances.

\begin{figure}[H]
    \centering
    \includegraphics[width=.7\textwidth]{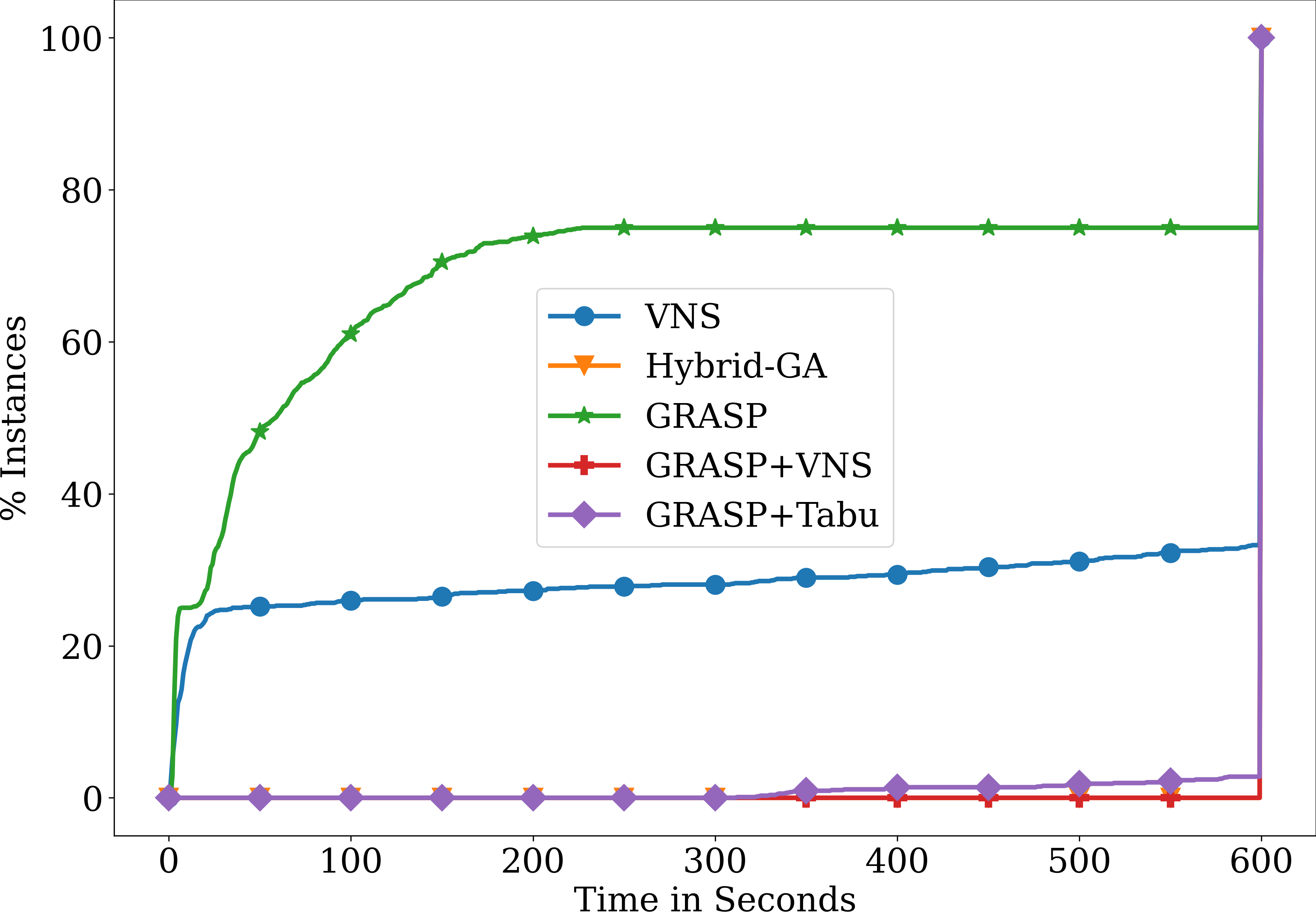}
    \caption{Graph of time for MAXSPACE-RDWV.}
    \label{fig:resultados_MAX-RDWV_tempo}
\end{figure}

In Figure~\ref{fig:resultados_MAX-RDWV_tempo}, we present a graph with the time comparison of the algorithms implemented for MAXSPACE-RDWV\@. We can see that GRASP is the least time-consuming algorithm and can finish execution for approximately~$75\%$ of instances within~$200$ seconds. Hybrid-GA, GRASP$\plus$Tabu and GRASP$\plus$VNS are the most time-consuming algorithms, reaching a time limit of~$600$ seconds in approximately~$100\%$ of instances.

Based on the graphs present before, we considered VNS and GRASP$\plus$VNS as our best algorithms for MAXSPACE-RDWV\@. In Table~\ref{tab:comp_rdwv}, we present a comparison of solutions value among these and Hybrid-GA for MAXSPACE-RDWV\@. Each cell of this table presents how many instances the algorithm of a line found a solution better than the solution found by the algorithm of a column. 

\begin{table}[ht]
\centering
\caption{Comparison of algorithms solutions for MAXSPACE-RDWV.}\label{tab:comp_rdwv}
\begin{tabular*}{\textwidth}{c @{\extracolsep{\fill}}ccc}
\toprule
& \textbf{Hybrid-GA} & \textbf{VNS} & \textbf{GRASP$\plus$VNS} \\ \midrule
\multicolumn{1}{c}{\textbf{Hybrid-GA}}  & - & 286  & 216 \\ 
\multicolumn{1}{c}{\textbf{VNS}} & 794  & - & 489  \\ 
\multicolumn{1}{c}{\textbf{GRASP$\plus$VNS}}  & 863 & 436 & - \\ 
\bottomrule
\end{tabular*}
\end{table}

In Table~\ref{tab:comp_rdwv}, we observe that GRASP$\plus$VNS is the algorithm that obtains the best solutions for MAXSPACE-RDWV, with~$863$ better solutions than Hybrid-GA and~$436$ better solutions than VNS\@.

\subsection{Statistical Analysis}

We performed a statistical analysis as presented by \citet{demvsar2006statistical} to compare the results obtained, identify a statistical difference between the proposed heuristics and verify if they are statistically better than the Hybrid-GA algorithm. For this, we use the \texttt{scmamp} and \texttt{stats} libraries of the \texttt{R} language.

We apply Friedman's test to show that the algorithms differ statistically from each other. The \textit{p-value} obtained by the Friedman test was less than $0.5$ for both MAXSPACE and MAXSPACE-RDWV\@. This means that, in both problems, there are statistical differences between the results obtained by the algorithms.

Thus, we apply a \textit{post-hoc} test to find which algorithms differ from each other. We use the Nemenyi test to compute the critical difference and identify groups of statistically equivalent algorithms. We say that the rank of an algorithm is~$1$ for a given instance if it finds the best solution for that instance among the considered algorithms, an algorithm is rank~$2$ if it obtains the second-best solution, and so on. The average rank of an algorithm is the average of the ranks obtained by that algorithm for a set of instances. In this test, the average ranks of the algorithms and the value of the critical difference (CD) are calculated.

In the following, we show a graphical representation of the results of this test. This representation distributes the algorithms from left to right in rank order (from lowest to highest). There is a horizontal bar connecting two algorithms when their average ranks differ by at most the value of CD, i.e., they are statistically equivalent.

\begin{figure}[H]
    \centering
    \includegraphics[width=.7\textwidth]{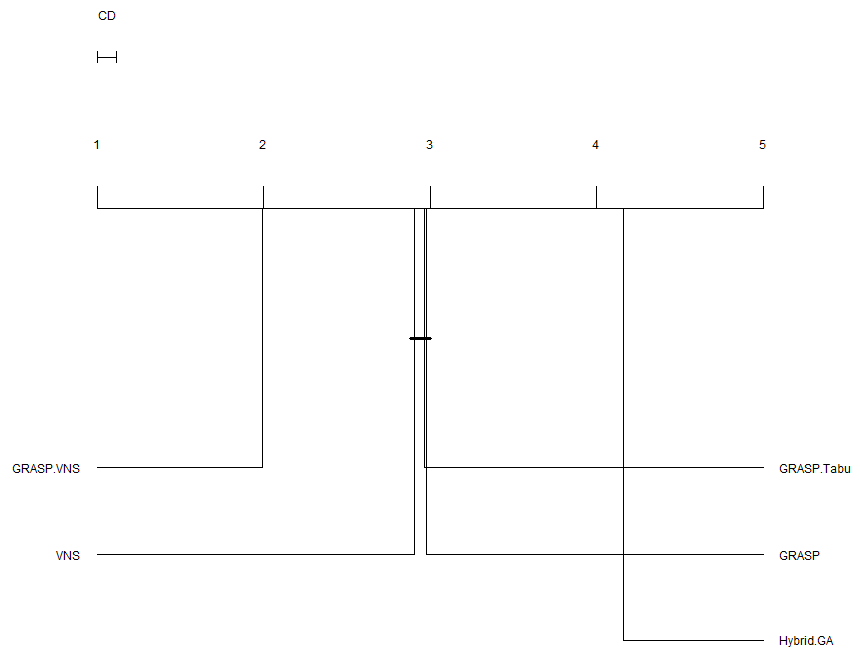}
    \caption{Nemenyi test for MAXSPACE.\label{fig:nem_MAX}}
\end{figure}

Figure~\ref{fig:nem_MAX} shows the result of the Nemenyi test for MAXSPACE\@. The GRASP$\plus$VNS was the lowest average rank algorithm for this problem, and the heuristics GRASP$\plus$Tabu, GRASP, and VNS obtained statistically equivalent results. Also, note that the worst average rank was from the Hybrid-GA algorithm.

\begin{figure}[H]
    \centering
    \includegraphics[width=.7\textwidth]{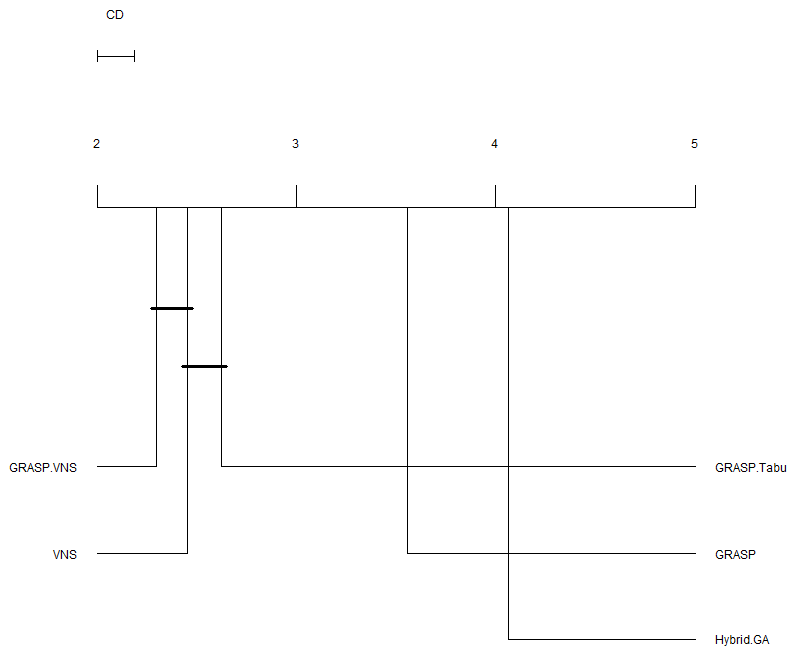}
    \caption{Nemenyi test for MAXSPACE-RDWV.\label{fig:nem_MAX_RDWV}}
\end{figure}

Figure~\ref{fig:nem_MAX_RDWV} presents the result of the Nemenyi test for the MAXSPACE-RDWV\@. As in MAXSPACE, the GRASP$\plus$VNS was the algorithm with the lowest average rank, but it obtained results statistically equivalent to those of the GRASP$\plus$Tabu and VNS heuristics. Again, the worst average rank is from the Hybrid-GA algorithm.

Thus, we conclude that, statistically, our best heuristic is GRASP$\plus$VNS and that there are different statistics between it and the Hybrid-GA algorithm.


